\documentstyle[12pt]{article}

\def\al{\alpha}

\def\be{\begin{equation}}
\def\ee{\end{equation}}

\begin{document}

\title{Dynamical supersymmetry breaking and unification of couplings}
\author{{\em S.~L.~Dubovsky\footnote{{\em e-mail:} sergd@ms2.inr.ac.ru}, 
             D.~S.~Gorbunov\footnote{{\em e-mail:} gorby@ms2.inr.ac.ru},
         and S.~V.~Troitsky\footnote{{\em e-mail:}    st@ms2.inr.ac.ru}}\\
   {\small {\em Institute for Nuclear Research of the 
  Russian Academy of Sciences,}}\\
  {\small {\em 60th October Anniversary 
  prospect, 7a, Moscow 117312, Russia.}}
  }
\date{July 15, 1997}
\maketitle
\begin{abstract}
We consider the possibility of unification of the 
Supersymmetric Standard Model
gauge groups with those of the dynamical supersymmetry 
breaking (DSB) sector in theories with gauge mediated
supersymmetry breaking.
We find constraints on the DSB gauge group beta function 
that come from unification of the gauge coupling constants
of the two sectors. These constraints are satisfied by a
fairly  wide class of models. We discuss possible
 unification scenarios in the context of a simple model.
\end{abstract}

{\bf 1.} Several mechanisms have been suggested to explain
supersymmetry breaking which should be present in the Minimal
Supersymmetric Standard Model, MSSM (for a recent review, see, 
e.g., Ref.\cite{kazakov}). One of them assumes that supersymmetry is
broken at very high energies in a hidden sector which interacts
with the MSSM fields only by gravity (see Ref.\cite{nilles} for a
review). Another possibility is that the dynamical supersymmetry
breaking (DSB) occurs at relatively low energies (of order $10^5$ to
$10^8$ GeV), again in a new sector, and is then transmitted to MSSM by
chiral superfields that interact with both sectors (see
Refs. \cite{newtools,Dine} and references therein). In this
paper we consider the latter class of theories where supersymmetry
breaking in MSSM has nothing to do with gravity.

One of the most prominent features of the MSSM is the unification of
electroweak and strong coupling constants at the scale $M_{GUT}\sim
10^{16}$ GeV. This is a strong indication to  Grand Unification. It is
interesting to consider whether it is possible that the gauge coupling
constants of the MSSM are unified with those of the DSB sector. This would
be a prerequisite for unification of both sectors into a single Grand
Unified Theory based on a simple gauge group. 

{\bf 2}.
In the DSB scenarios, strong interactions in the new sector force 
$F$-components of several fields to acquire vacuum expectation values.
They are determined by the scale $\Lambda$ at which the coupling
constant of the gauge group of the DSB sector becomes strong.
In fact, the DSB models often involve product gauge groups. In that case
the dynamics of supersymmetry breaking is usually driven by the 
strongest coupling, i.e., the $F$-components are determined by the largest 
value of the infrared pole $\Lambda$.
In most low energy supersymmetry breaking models, this effect
is fed down to the MSSM by  several fields interacting
with both sectors. Soft terms in the MSSM are generated by loop effects, 
so the supersymmetry breaking scale in the MSSM is lower than $\Lambda$
by factors involving strong and/or electroweak coupling constants. 
The actual value of the DSB scale $\Lambda$ depends strongly
on the mechanism of transmission of nonzero $F$'s to soft terms in
the MSSM, so the estimates of $\Lambda$ vary in different
models
\cite{Dine1,Dine2,newtools}. We allow
$\Lambda$ to take values between $10^5~{\rm GeV}$  and $10^8~{\rm GeV}$. 
The lower bound comes from the requirement that superpartners
of ordinary particles are not too light, while the upper one
is implied by the bounds on gravitino mass and by the analysis of
nucleosynthesis \cite{dimop}.
 
{\bf 3.}  Several scenarios of unification of gauge couplings of both
sectors can be suggested.  The first possibility (Fig.~1) is that the DSB
gauge coupling is unified with the ordinary GUT coupling somewhere
between $M_{GUT}$ and the fundamental scale, $M_s\sim 10^{18}~ {\rm
GeV}$, where one believes that string effects become essential.
Alternatively, in case of product gauge group in the DSB sector, one may
suppose that unification in the MSSM and DSB sectors occurs
independently at $M_{GUT}$ and at some scale $m$, respectively, while
the couplings of both GUTs are unified at $M_s$ (Fig.~2). We begin with the
former scenario, as it does not require the knowledge of a concrete
mechanism of unification in the DSB sector.
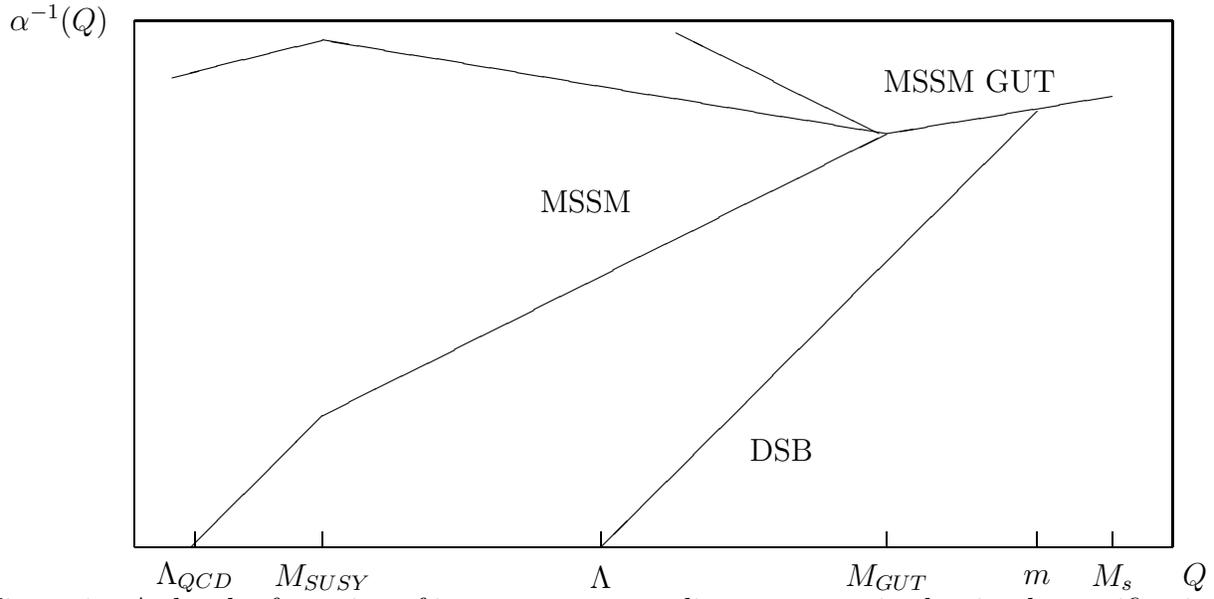
\begin{figure}
\unitlength=1.00mm
\special{em:linewidth 0.4pt}
\linethickness{0.4pt}
\begin{picture}(121.00,70.00)(-20.0,0.0)
\put(130.00,60.00){\line(-6,-1){30.00}}
\put(100.00,55.00){\line(-6,1){75.00}}
\put(25.00,67.50){\line(-4,-1){20.00}}
\put(99.00,55.00){\line(-2,1){27.00}}
\put(111.00,62.00){\makebox(0,0)[cc]{MSSM GUT}}
\put(86.00,13.00){\makebox(0,0)[cc]{DSB}}
\put(60.00,46.00){\makebox(0,0)[cc]{MSSM}}
\put(130.00,2.00){\line(0,-1){2.00}}
\put(120.00,0.00){\line(0,1){2.00}}
\put(100.00,2.00){\line(0,-1){2.00}}
\put(8.00,2.00){\line(0,-1){2.00}}
\put(25.00,2.00){\line(0,-1){2.00}}
\put(138.00,0.00){\line(0,1){70.00}}
\put(138.00,70.00){\line(-1,0){138.00}}
\put(0.00,70.00){\line(0,-1){70.00}}
\put(130.00,-4.00){\makebox(0,0)[cc]{$M_s$}}
\put(120.00,-4.00){\makebox(0,0)[cc]{$m$}}
\put(100.00,-4.00){\makebox(0,0)[cc]{$M_{GUT}$}}
\put(62.00,-4.00){\makebox(0,0)[cc]{$\Lambda$}}
\put(8.00,-4.00){\makebox(0,0)[cc]{$\Lambda_{QCD}$}}
\put(25.00,-4.00){\makebox(0,0)[cc]{$M_{SUSY}$}}
\put(7.50,0.00){\line(1,1){17.50}}
\put(25.00,17.50){\line(2,1){75.00}}
\put(141.00,-4.00){\makebox(0,0)[cc]{$Q$}}
\put(-10.00,70.00){\makebox(0,0)[cc]{$\alpha^{-1}(Q)$}}
\put(120.00,58.00){\line(-1,-1){58.00}}
\put(0.00,0.00){\line(1,0){138.00}}
\put(62.00,2.00){\line(0,-1){2.00}}
\end{picture}
\caption{
A sketch of running of inverse gauge
coupling constants in the simplest unification scenario
}
\end{figure}
\begin{figure}
\unitlength=1mm
\special{em:linewidth 0.4pt}
\linethickness{0.4pt}
\begin{picture}(143.00,70.00)(-20.0,0.0)
\put(130.00,60.00){\line(-2,-3){10.00}}
\put(120.00,45.00){\line(-4,-3){60.00}}
\put(120.00,45.00){\line(-2,-1){90.00}}
\put(130.00,60.00){\line(-6,-1){30.00}}
\put(100.00,55.00){\line(-6,1){75.00}}
\put(25.00,67.50){\line(-4,-1){20.00}}
\put(99.00,55.00){\line(-2,1){27.00}}
\put(129.00,48.00){\makebox(0,0)[cc]{DSB}}
\put(129.00,43.00){\makebox(0,0)[cc]{GUT}}
\put(111.00,62.00){\makebox(0,0)[cc]{MSSM GUT}}
\put(86.00,13.00){\makebox(0,0)[cc]{DSB}}
\put(60.00,46.00){\makebox(0,0)[cc]{MSSM}}
\put(130.00,2.00){\line(0,-1){2.00}}
\put(120.00,0.00){\line(0,1){2.00}}
\put(100.00,2.00){\line(0,-1){2.00}}
\put(60.00,2.00){\line(0,-1){2.00}}
\put(8.00,2.00){\line(0,-1){2.00}}
\put(25.00,2.00){\line(0,-1){2.00}}
\put(0.00,0.00){\line(1,0){138.00}}
\put(138.00,0.00){\line(0,1){70.00}}
\put(138.00,70.00){\line(-1,0){138.00}}
\put(0.00,70.00){\line(0,-1){70.00}}
\put(130.00,-4.00){\makebox(0,0)[cc]{$M_s$}}
\put(120.00,-4.00){\makebox(0,0)[cc]{$m$}}
\put(100.00,-4.00){\makebox(0,0)[cc]{$M_{GUT}$}}
\put(60.00,-4.00){\makebox(0,0)[cc]{$\Lambda$}}
\put(8.00,-4.00){\makebox(0,0)[cc]{$\Lambda_{QCD}$}}
\put(25.00,-4.00){\makebox(0,0)[cc]{$M_{SUSY}$}}
\put(7.50,0.00){\line(1,1){17.50}}
\put(25.00,17.50){\line(2,1){75.00}}
\put(141.00,-4.00){\makebox(0,0)[cc]{$Q$}}
\put(-10.00,70.00){\makebox(0,0)[cc]{$\alpha^{-1}(Q)$}}
\end{picture}
\caption{
Same as Fig. 1 but in the scenario with unification
in the DSB sector
}
\end{figure}
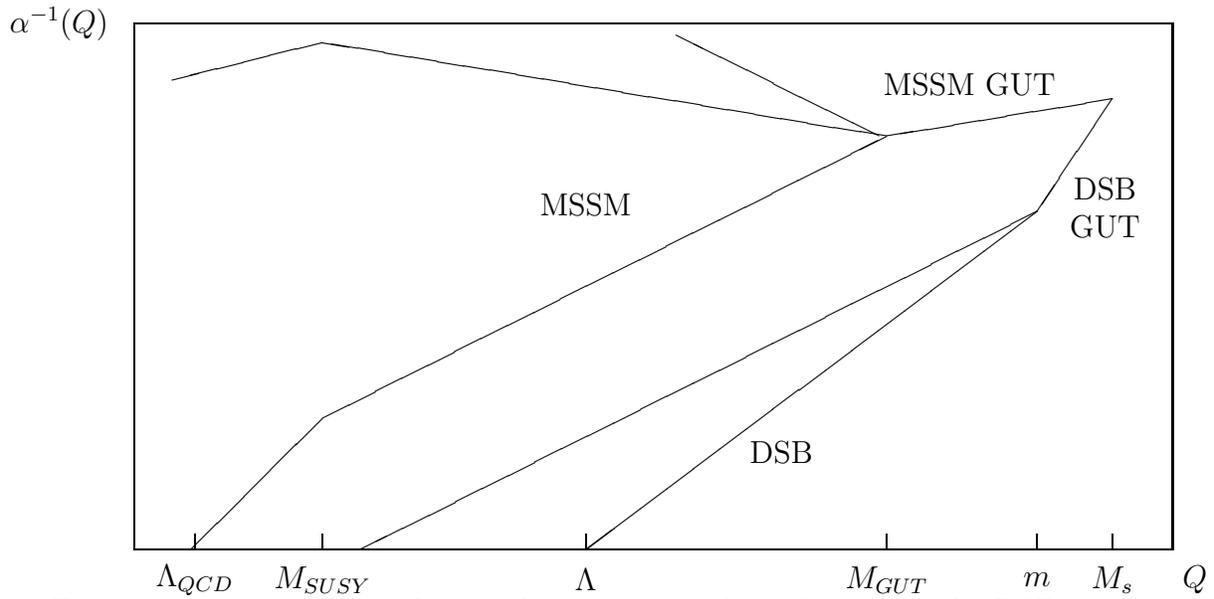

Let us assume for definiteness that the underlying GUT for the MSSM is
the minimal supersymmetric $SU(5)$ \cite{GUT}. Let $\al_G$ and $\al$
be the coupling constants of this $SU(5)$ and the DSB gauge group,
respectively.  In most of gauge mediation scenarios, the MSSM is
extended by adding messenger fields which fall into a vector-like
representation of $SU(5)$.  We assume that they belong to a single
$(5+\bar{5})$ representation, so the gauge couplings of the MSSM group
$SU(3)\times SU(2)\times U(1)$ unify at $M_{GUT} = 10^{15.8}~{\rm
GeV}$ 
\cite{kazakov} to a value of $ \al_G(M_{GUT})\approx 1/23 $ (we
use $\alpha_S(M_Z)=0.118$ and the thresholds $M_{SUSY}=300~{\rm GeV}$
for SUSY particles and $M=10^{5}~{\rm GeV}$ for
messengers; in fact our results are practically insensitive to the
choice of values of $M_{SUSY}$ and $M$ within the range outlined 
above).  
The unification condition in the scenario shown in Fig.1 is
$$
\al_G(m)=\al(m), ~~~~~ M_{GUT} <m<M_s.
$$
We use the one loop beta function for supersymmetric gauge theories 
with matter,
$$
\beta(\alpha)=-{\alpha^2 \over 2\pi} C,
~~~~
C= \left(3 T(G)-\sum_i T(R_i)\right),
$$
where $T(G)$ and $T(R_i)$ are Dynkin indices (one half of 
quadratic Casimir operators) for adjoint representation and for the
$i$-th matter superfield representation of the gauge group,
respectively. For $SU(N)$ group, one has $T(G)=N$, 
and  each fundamental or
anti-fundamental matter multiplet contributes $T(R_i)=1/2$. 
For the ordinary supersymmetric $SU(5)$ GUT 
$C=2$ with messenger fields taken into account.

Let us suppose first that in the DSB sector, there are no fields with
thresholds between $\Lambda$ and $M_s$. It is straightforward
to see that the restriction $10^{5}~{\rm GeV}<\Lambda<10^{8}~{\rm GeV}$ 
places bounds on the beta function for the
gauge coupling with largest infrared pole in the DSB sector
\be
4.6<C<7.9.
\label{c}
\ee
Among relatively simple models exhibiting DSB (see Refs.
\cite{Dine,Skiba,newtools} for reviews), only two 
have beta functions satisfying (\ref{c}).  The first
one is the well known``3-2'' model \cite{ADS}, one of the simplest
models explored from the point of view of the DSB scenario \cite{ADS,Dine2},
which has $C=7$. Another one is based on the same group $SU(3)\times
SU(2)$ but
with somewhat more complicated matter content that consists of
the following superfields (numbers in parenthesises indicate $SU(3)\times
SU(2)$ representations):
\be
\begin{array}{c} 
{\rm one~ field}~ Q~ (3,2),\\
{\rm three~ fields} ~\bar{L}_I~(\bar{3},1),~I=1,2,3,\\
{\rm one~ field} ~L~(3,1),\\
{\rm three~ fields} ~\bar{R}_A~(1,\bar{2}),~A=1,2,3.\\
\end{array}
\label{content}
\ee
This model is a
representative of a series of theories based on product groups
$SU(N)\times SU(N-M)$ in which DSB occurs through dynamics of the dual
group \cite{IntrThom,product1,product2} 
upon adding the superpotential of the form 
\be 
W=\lambda^{IA}\bar{L}_I Q \bar{R}_A + {\cal{M}}^I \bar{L}_I L.
\label{W}
\ee
In Eq.\ (\ref{W}), $\lambda^{IA}$ is the matrix of Yukawa constants 
of rank 2 or 3, and ${\cal{M}}$ is a $3\times 1$ mass matrix.
In this model $C=6$ (in both cases it is the $SU(3)$ gauge interactions
that 
drive the dynamics of supersymmetry breaking).

Let us now turn to the case when additional heavy matter 
superfields with
mass of order $M_x\gg\Lambda$ are present in the DSB model. They do not 
affect
the low-energy dynamics responsible for supersymmetry 
breaking since they 
can be integrated out from the effective action. However, new threshold
appears at $M_x$, and  the first coefficient of the beta function
becomes smaller at scales greater than $M_x$. 
So, if sufficiently heavy 
matter 
superfields are introduced, any model which initially had $C>4.6$ becomes
acceptable. In this way almost all known 
DSB models can be made consistent
with the unification of the MSSM and DSB gauge couplings.

 {\bf 4.}  It is worth noting that the model with matter
content (\ref{content}) is actually unifiable to an $SU(5)$ gauge
theory broken down to $SU(3)\times SU(2)\times U(1)$ by a
Higgs 24-plet, in  the same manner as in the 
$SU(5)$  GUT of the Standard 
Model \cite{GUT}.  Indeed, consider breaking  
$SU(5) \to SU(3)\times SU(2)$ in the DSB 
sector and add following matter multiplets: \be
\begin{array}{c}
{\rm one~field}~10~=~(3,2)+(\bar{3},1)+(1,1),\\
{\rm one~field}~5~=~(3,1)+(1,2),\\
{\rm two~fields}~\bar{5}_{i}~=~(\bar{3},1)_i+(1,\bar{2})_i,~i=1,2.
\end{array}
\label{su5content}
\ee
Due to the equivalence between the fundamental and conjugate
representations of $SU(2)$, it has exactly the matter content
(\ref{content}) -- up to a singlet which does not take part in
$SU(3)\times SU(2)$ dynamics (but carries a $U(1)$ charge). 
The required superpotential (\ref{W}) can be written it terms of these
$SU(5)$ multiplets,
$$
W=\lambda 10_{\alpha\beta} \bar{5}^\alpha_1 \bar{5}^\beta_2 +
  m 5_\alpha \bar{5}^\alpha_1.
$$
This superpotential includes couplings required for DSB, Eq.\
(\ref{W}), as well as some additional terms which do not spoil the
mechanism of supersymmetry breaking.  In this way we arrive at
a model with the gauge symmetry $SU(5)_{SM}\times SU(5)_{DSB}$.

Within this example of unification in the
DSB sector, we may
discuss a possibility that the two $SU(5)$'s
unify at the fundamental scale $M_s$ (and each breaks down to
$SU(3)\times SU(2)\times U(1)$, at $M_{GUT}$ and $m$ in
the MSSM and
DSB sectors, respectively). This is an example of the scenario 
illustrated in 
Fig.2. Mass 
scale $m$ is now a free parameter satisfying \be
\Lambda<m<M_s.
\label{selfc}
\ee
We use this restriction together with the unification
constraints for coupling constants to find that the picture is
self-consistent at 
$10^{6.6}~{\rm GeV}<\Lambda <10^{8}~{\rm GeV}$, if
constraints from  nucleosynthesis are taken into account.

In this scenario, mass scales $m$ and $M_{GUT}$ are introduced by
hand. It would be natural for {\em all} groups to unify at $M_s$; in the
case of the MSSM this would require more complicated 
messenger sector whose structure can hardly be anticipated
without concrete mediation mechanism being favored.
The appealing possibility of a single unification scale $M_s$
may be realized in
a Grand Unified Theory which contains both MSSM and DSB sectors and has
sufficiently large gauge group and matter content. The above discussion
points to 
$SU(5)_{SM}\times SU(5)_{DSB}$;
the DSB sector may even have the same matter content as the MSSM
if two flavors are made heavy by some mechanism.

The matter content of the ``3-2'' model unifies in $SU(5)$ multiplets
10 and $\bar{5}$. Since no invariant superpotential can be written for
$SU(5)$ with single 10 and $\bar{5}$ while tree level superpotential for
$SU(3)\times SU(2)$ is required for DSB, the superpotential
 should be generated by
some additional mechanism after breaking of $SU(5)$. Even in
this case the model does not satisfy the requirements for the unification
within the scenario sketched in Fig.~2 since bounds on $\Lambda$ imposed
by nucleosynthesis are inconsistent with (\ref{selfc}) in this model.
                    
Recently, the problem of unification of the MSSM and DSB gauge coupling
constants has been discussed in Ref.\ \cite{k} with the results
opposite to ours. We believe 
that 
the difference in restrictions on the beta function coefficient between
Eq.(\ref{c}) and Ref.\ \cite{k} is due to a numerical error
in Ref.\ \cite{k}. Also, the possibility of additional heavy thresholds
has not been considered in Ref.\ \cite{k}.

{\bf 5.}  To summarize, we have found 
constraints on the content of the DSB
sector 
under which its gauge couplings are unified with those of the MSSM and
its GUT.
In the minimal version when no additional heavy matter superfields
exist in the DSB sector, these bounds choose two models out of a 
variety of 
theories exhibiting DSB. 
The model with $SU(3)\times SU(2)$ and matter content (\ref{content}) is
unifiable to $SU(5)$ so that $SU(5)\times
SU(5)$ gauge group is favored. 
Almost every known DSB model can be made 
consistent with gauge
unification by choosing appropriate set of
heavy chiral superfields  in the DSB sector with 
threshold $M_x\gg\Lambda$. 

 The authors are indebted to M.V.Libanov and
V.A.Rubakov for numerous helpful discussions.  This work is supported
in part by Russian Foundation for Basic Research grant
96-02-17449a. The work of S.T. is supported in part by the U.S.\
Civilian Research and Development Foundation for Independent States of
FSU (CRDF) Award No.~RP1-187, by INTAS grant 94-2352, 
and by ISSEP fellowship, grant
No.~a97-45.

\end{document}